\documentclass[%
 reprint,
 amsmath,amssymb,
 aps,
 prl
]{revtex4-1}

\usepackage{graphicx}
\usepackage{dcolumn}
\usepackage{bm}
\usepackage{hyperref}
\usepackage[mathlines]{lineno}

\usepackage{xcolor}
\begin{document}
\preprint{arXiv}

\title{
  Renormalized perturbation theory at large expansion orders
}
\author{Riccardo Rossi$^1$,
  Fedor \v{S}imkovic$^{2,3}$ IV,
  and Michel Ferrero$^{2,3}$}

\affiliation{$^1$Center for Computational Quantum Physics, Flatiron Institute, 162 5th Avenue, New York, NY 10010 \\
$^2$CPHT, CNRS, Ecole Polytechnique, Institut Polytechnique de Paris, Route de Saclay, 91128 Palaiseau, France\\
$^3$Coll\`ege de France, 11 place Marcelin Berthelot, 75005 Paris, France}

\date{\today}

\begin{abstract}
  We present a general formalism that allows for the computation of large-order renormalized expansions in the spacetime representation, effectively doubling the numerically attainable perturbation order of renormalized Feynman diagrams. We show that this formulation compares advantageously to the currently standard techniques due to its high efficiency, simplicity, and broad range of applicability. Our formalism permits to easily complement perturbation theory with non-perturbative information, which we illustrate by implementing expansions renormalized by the addition of a gap or the inclusion of Dynamical Mean-Field Theory. As a result, we present numerically-exact results for the square-lattice Fermi-Hubbard model in the low temperature non-Fermi-liquid regime and show the momentum-dependent suppression of fermionic excitations in the antinodal region.
\end{abstract}

\maketitle

Renormalization is one of the most fruitful ideas in physics. Originally discovered as a method to eliminate one-loop Feynman-diagram infinities in quantum electrodynamics (QED)~\cite{qed_bethe, qed_tomonaga, qed_feynman, qed_schwinger}, it has notably lead to one of the most precise comparison of theory and experiment~\cite{qed_exp_th}. The usefulness of renormalization beyond high-energy physics has soon been understood and it was applied to condensed matter physics~\cite{anderson_rg, anderson_poor} and especially to the theory of critical phenomena~\cite{wilson_epsilon_expansion}, which has led to the development of the perturbative renormalization group technique. As computing renormalized Feynman diagrams is at the core of our quantitative understanding of nature, it is of critical importance to find efficient strategies in order to successfully perform computations. Yet, evaluating a large number of diagram orders is an extremely challenging task even with modern computational facilities. For example, no 6-loop QED computation has been attempted to date, despite strong interest due to the availability of high-precision experiments. The main limitation in computing large-order contributions is the ''factorial barrier`` represented by the factorial growth of the number of Feynman diagrams with increasing diagram order.
The Diagrammatic Monte Carlo algorithm~\cite{ProkSvistFrohlichPolaron,ProkofevSvistunovPolaronLong,VanHoucke1short, kris_felix, deng} was introduced with the idea that Feynman diagrams could be good sampling variables in strongly-correlated electronic systems as they can be defined directly in the thermodynamic and continuum limit.  While this approach dramatically simplified and automatized the computation of Feynman diagrams, and there are many recent advancements in this direction~\cite{wu_controlling, kun_chen, gull_inchworm, vucicevic2019real, taheridehkordi2019algorithmic, taheridehkordi2019optimal}, it is still limited by the factorial barrier as it considers explicit Feynman diagram topologies.

A recently introduced algorithm overcomes the factorial barrier by computing connected and irreducible bare Feynman diagrams~\cite{cdet, alice_michel, fedor_sigma, rr_sigma} at a computational cost growing exponentially with expansion order, which allowed for the computation of an unprecedented number of Feynman-diagram orders directly in the thermodynamic limit. Similarly effective exponential algorithms overcoming the factorial barrier have also been found for the real-time evolution of quantum systems~\cite{olivier, corentin, kid_gull_cohen, moutenet2019cancellation}, allowing to reach the large-time limit in quantum impurity models.

While bare expansions are remarkably powerful for fermionic systems on a lattice and at finite temperatures~\cite{wu_controlling, fedor_sigma, fedor_hf, kim_cdet}, given the finite radius of convergence and the resulting polynomial complexity of the many-body problem~\cite{rr_epl}, they are nevertheless limited by the appearance of poles in the complex plane. These poles may prohibit the resummation of the series in vicinity of sharp crossovers~\cite{fedor_sigma}. Moreover, in the very-low temperature regime infrared divergencies are generically present~\cite{feldman}. Renormalization is the fundamental missing tool to continue to make progress. For example, it has been shown that optimized chemical potential shifts can already yield drastic improvements to the properties of evaluated series~\cite{rubtsov2005continuous, wu_controlling}. Furthermore, when considering systems directly in the continuous space, one is generally forced to perform renormalization from the start in order to be able to even define the theory.



In this Letter, we present a general formalism that allows for the numerical computation of renormalized perturbative expansions at large expansion orders. More precisely, we prove that it is possible to overcome the factorial barrier: We compute factorially-many renormalized Feynman diagrams in the spacetime representation with only an exponential cost, independently of the renormalization procedure. In what follows, we introduce the underlying theoretical concepts and show examples of large order computations, up to $10-14$ orders, for the square-lattice Fermi-Hubbard model in non-perturbative regimes, where no other controlled techniques are applicable. To achieve this goal we have designed new renormalization schemes using non-perturbative information, such as approximate solutions from Dynamical Mean-Field Theory (DMFT~\cite{antoine_dmft}), altering the bare series in a way that extends the convergence radius and the applicability of perturbation theory. Finally, we show that the renormalized expansion can be used in the non-Fermi-liquid regime of the Hubbard model by computing the spectral function, which shows a strong suppression of antinodal quasiparticles.


\begin{figure}
  \includegraphics[width=0.45\textwidth]{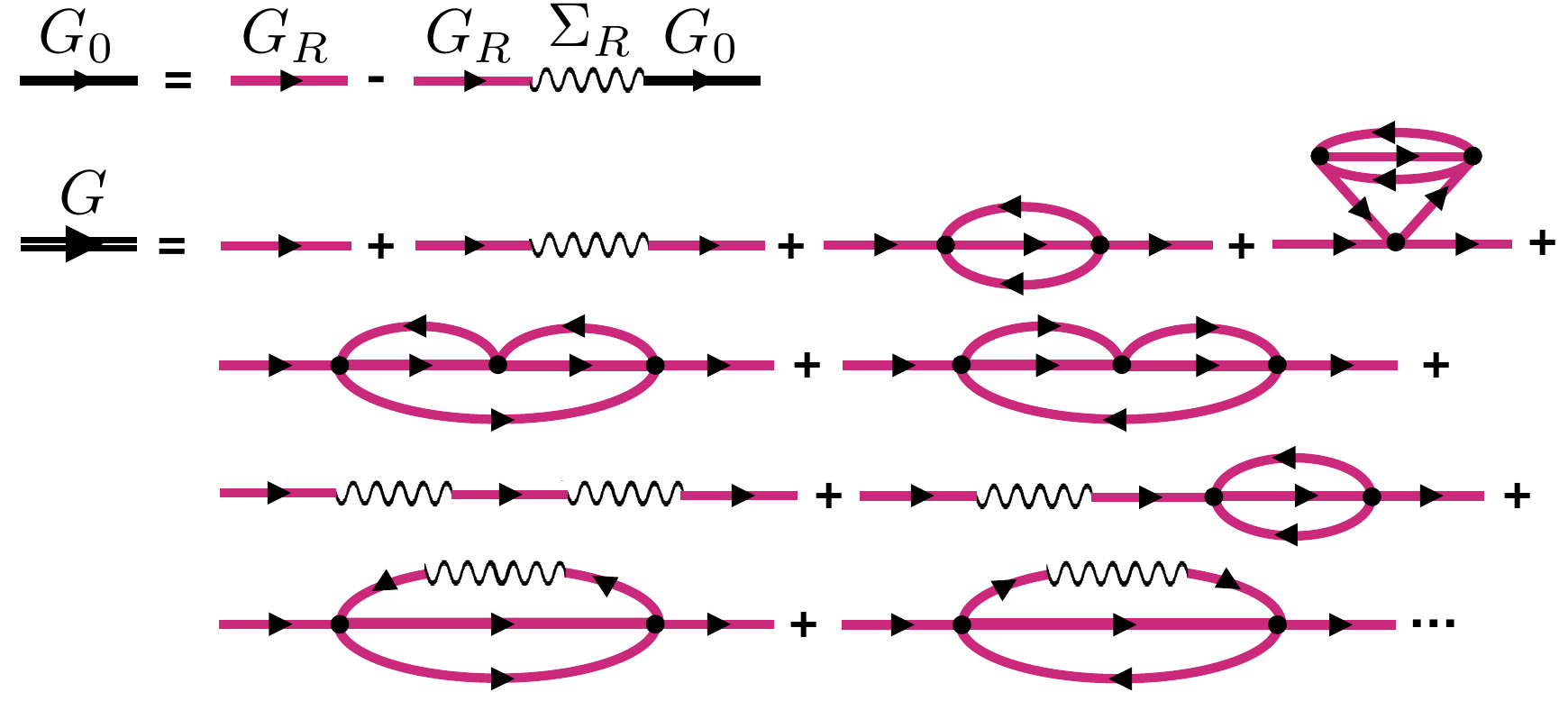}
\caption{Quadratic self-energy renormalization. Top: Dyson-like equation relating the bare Green's function $G_0[\xi]$, the renormalized Green's function $G_R$ and the self-energy functional $\Sigma_R[\xi]$. Bottom: explicit examples of diagrams contributing to the physical Green's function $G[\xi]$. \label{Fig_quadratic_renormalization}}
\end{figure}

For concreteness, we focus our discussion on one-particle renormalizations for the Hubbard model in the following theoretical part. We introduce a generalization of the ``shifted action'' of Ref.~\cite{shifted_action}
\begin{equation}
S_{\text{bare}}[G_0,\xi]=-\langle\varphi| G_0^{-1}|\varphi\rangle+\int_X \xi(X)\,\left(\bar{\varphi}_\uparrow\bar{\varphi}_\downarrow \varphi_\downarrow\varphi_\uparrow\right)(X),
\end{equation}
where $\varphi_\sigma$ is a Grassmann field, $X$ is the imaginary-time-lattice coordinate, and $\xi(X)$ is a spacetime dependent coupling constant. The expansion in powers of $\xi$ reproduces exactly the bare expansion in the spacetime representation. For example, the Green's function
\begin{equation}
G[\xi](Y,Y'):=-\frac{\int e^{-S_{\text{bare}}[G_0,\xi]}\;\varphi(Y)\,\bar{\varphi}(Y')}{\int e^{-S_{\text{bare}}[G_0,\xi]}}
\end{equation}
can be written as
\begin{equation}\label{power_expansion}
G[\xi](Y,Y')=\sum_{n=0}^\infty\frac{1}{n!}\int_{1,\dots,n}\hspace{-5mm}G_{Y,Y'}(\{X_1,\dots,X_n\})\prod_{j=1}^n\xi(X_j),
\end{equation}
where the functional derivative with respect to $\xi(X_j)$, $G_{Y,Y'}(\{X_1,\dots,X_n\})$, is the sum of all connected bare Feynman diagrams with $X_1\dots X_n$ as internal vertex positions, symmetrized with respect to permutations of the internal vertices, and with $Y$ and $Y'$ as external points. One-particle renormalization in the spacetime representation can be achieved by substituting $G_0$ with a functional of the interaction $\xi$, which we denote by $\tilde{G}_0[\xi]$ 
\begin{equation}
\left(\tilde{G}_0[\xi]\right)^{-1} =: G_R^{-1} + \Sigma_{R}[\xi].
\end{equation}
The functional $\tilde{G}_{0}[\xi]$ is equal to an arbitrary $G_R$ at zero interaction, and it coincides with the physical bare propagator $G_0$ for $\xi(X)=U$, at the value of the interaction strength $U$ we are interested in. There are no restrictions on the functional $\Sigma_{R}[\xi]$, in particular it can be defined as an infinite series in $\xi$ without computational overhead. We can now define the renormalized action functional as
\begin{equation}\label{renormalized_action}
S_R[G_R, \xi] := S_{\text{bare}}[G_0[\xi], \xi]
\end{equation}
and use it to define the Green's function as the $\xi$ functional
\begin{equation}
G[\xi](Y,Y'):=-\frac{\int e^{-S_{R}[G_R,\xi]}\;\varphi(Y)\,\bar{\varphi}(Y')}{\int e^{-S_{R}[G_R,\xi]}}.
\end{equation}
The sum of all renormalized Feynman diagrams for fixed symmetrized spacetime positions is then obtained by expanding in powers of $\xi$.  $G_{Y,Y'}(\{X_1,\dots,X_n\})$ from Eq.~\eqref{power_expansion} is now the sum of all symmetrized renormalized Feynman diagrams with internal vertex positions $X_1$, $\dots$, $X_n$.
For example, the fully-renormalized one-particle scheme, also called ``bold'', is obtained by imposing that $\Sigma_{R}[\xi]$ is such that $G[\xi](Y,Y')$
is a constant functional identically equal to $G_R$: $G[\xi]=G_R$. In this case, $\Sigma_{R}[\xi]$ can be diagrammatically constructed as the sum of all two-particle irreducible, ``bold'', diagrams. Not all renormalization schemes have a simple Feynman-diagrammatic interpretation: renormalization using Feynman diagrams becomes quickly unmanageable with order in the general case as one has to keep track of all possible counter-terms.


\begin{figure}
  \includegraphics[width=0.35\textwidth]{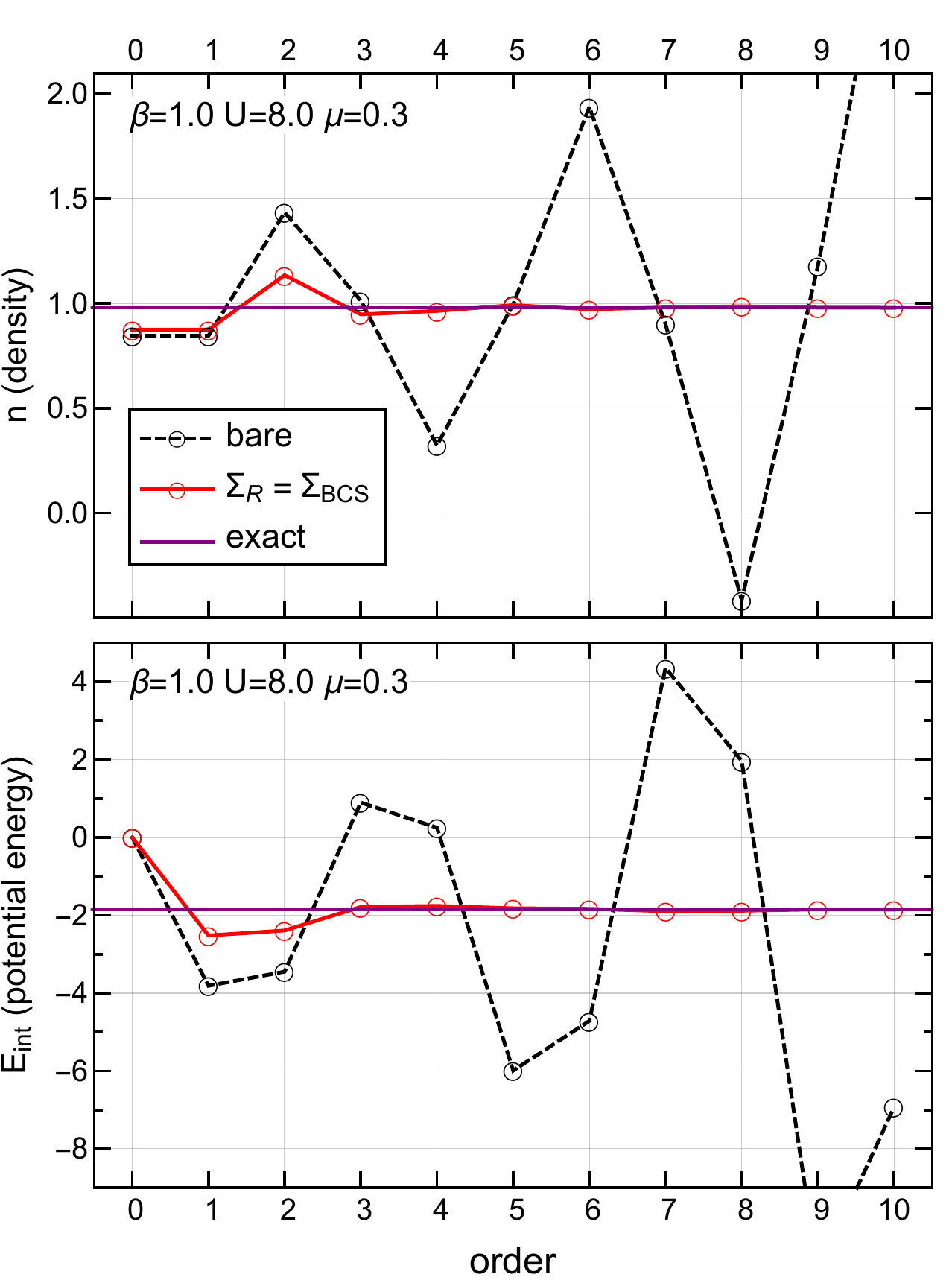}
\caption{Partial sums of the bare series (dashed black lines) and BCS-shifted series (red lines) for Hubbard atom density (top plot) and potential energy (bottom plot) are compared to analytic results (purple lines). \label{Fig_Hubbard_atom}}
\end{figure}

A key observation is that Eq.~\eqref{renormalized_action} implies that renormalized expansions are equivalent to bare expansions with a functional propagator $\tilde{G}_0[\xi]$. We can therefore apply the CDet algorithm~\cite{cdet} for bare expansions, provided we can generalize it to functionals. We introduce an efficient and general way to deal with functional expansions based on ``nilpotent polynomials'', functions of $n$ commuting symbols $z_j$ such that $z_j^2=0$, for $j\in\{1,\dots,n\}$. If we want to compute $n$-th order functional derivatives with respect to $\xi(X_1)\, \dots \,\xi(X_n)$, we only need to expand the functional up to linear order in each $\xi(X_j)=:z_j$, $j\in\{1,\dots,n\}$, discarding any term of order $z_j^2$ and higher. Nilpotent polynomials of $n$ variables form a ring, where multiplication is defined as a subset convolution, which can be performed in $O(3^n)$ operations, or alternatively $O(n^2\,2^n)$ operations using a fast subset convolution~\cite{koivisto}:
\begin{equation}
\begin{split}
&Q_3(z_1,\dots,z_n) = Q_1(z_1,\dots,z_n)\cdot Q_2(z_1,\dots,z_n) \nonumber\\
&\Longleftrightarrow \quad Q_3(V)=\sum_{S \subseteq V} Q_1(S)\,Q_2(V\setminus S),
\end{split}
\end{equation}
where  $Q_i(V)$ is the coefficient of $\prod_{j\in V} z_j$. Interestingly, the recursive formula from Ref.~\cite{cdet} which is used to compute sums of connected diagrams for the bare expansion based on $S_{\text{bare}}[G_0, \xi]$ can be reinterpreted as polynomial division between two nilpotent polynomials~\cite{rr_sigma}:
\begin{equation}
\begin{split}
&Q_3(z_1,\dots,z_n)=Q_1(z_1,\dots,z_n) / Q_2(z_1,\dots,z_n) \nonumber \\
&\Longleftrightarrow \quad Q_3(V) = Q_1(V) - \sum_{S\subsetneq V} Q_3(S)\,Q_2(V\setminus S),
\end{split}
\end{equation}
where $Q_2(\emptyset)=1$ is assumed. We can therefore obtain a fast algorithm for renormalized expansions by considering a nilpotent-polynomial-valued bare propagator $\tilde{G}_0(z_1,\dots,z_n)$ and use the CDet algorithm. For fermionic systems we need to compute the sum of determinants of nilpotent-polynomial matrices, a task that can be performed using additions, multiplications, and divisions of nilpotent polynomials. One can show that the computational cost to compute the sum of all symmetrized renormalized Feynman diagrams for fermionic systems for a given configuration of interaction vertices is $O(n^3 \,4^n)$, or alternatively $O(n^5 \,3^n)$ by using fast subset convolutions~\footnote{This comes from the fact that the computational cost is dominated by the computations of the $2^n$ determinants of nilpotent polynomials of variable order.}
, much better than $(n!)^2$ which is a lower bound on the cost of enumerating all diagrams over all permutations of internal vertices.

\begin{figure}
  \includegraphics[width=0.35\textwidth]{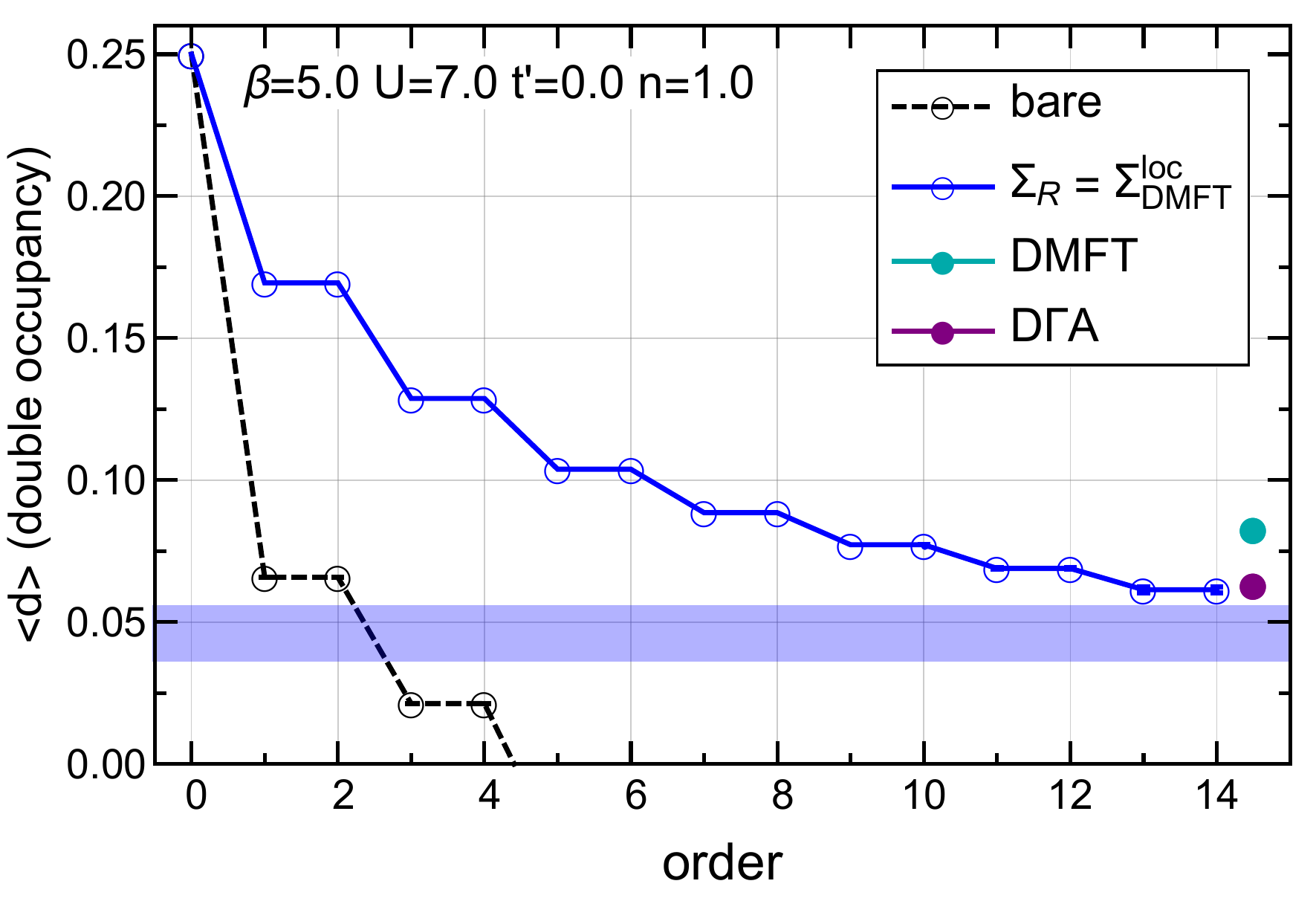}
\caption{Partial sums for the double occupancy for the half-filled Hubbard model. The bare series (dashed black lines) and the DMFT-shifted series (blue lines) and  the resummated result for the DMFT-shifted series (Pad\'e approximants, horizontal blue line) are shown. The double occupancy from DMFT (green) and a linear D$\Gamma$A extrapolation (purple) are plotted for comparison (data courtesy of T.~Sch\"afer). \label{Fig_double_occupancy}}
\end{figure}

We now focus our attention on the Fermi-Hubbard model on the square lattice, defined by the Hamiltonian
\begin{equation}
\hat{H}=-\sum_{i,j, \sigma}t_{ij}\,\hat{\psi}_{\sigma i}^\dagger \,\hat{\psi}^{\phantom{\dagger}}_{\sigma j} + U \sum_{i}\hat{\psi}_{\uparrow i}^\dagger \,\hat{\psi}_{\downarrow i}^\dagger\,\hat{\psi}^{\phantom{\dagger}}_{\downarrow i}\,\hat{\psi}^{\phantom{\dagger}}_{\uparrow i},
\end{equation}
where $i$, $j$ are lattice sites, $t_{ij}=1$ for nearest neighbors, $t_{ij}=t'$ for next-nearest neighbors, and zero otherwise. We are interested in the repulsive model where $U>0$ with an average number of particles per site $n$ close to one. In the regime of low temperatures and/or high interactions the bare expansion becomes very difficult to compute and evaluate, seemingly due to infrared divergencies coming from shifts of the non-interacting Fermi surface. Moreover, the presence of a superfluid instability in the attractive model ($U<0$) reduces the radius of convergence of the series to zero at zero temperature.

Two main renormalization approaches have been proposed to cure the bare expansion: The first is a fully self-consistent (bold) formalism which eliminates infrared divergencies by using the physical propagator in the expansion~\cite{deng}. However, a known problem with this approach is that the bold scheme can converge towards unphysical answers at strong interactions~\cite{kfg,rossi2015skeleton}. A second approach is a renormalized perturbation theory at fixed Fermi surface~\cite{feldman}, but has the caveat that it supposes the actual existence of a Fermi surface, which can be destroyed at strong interactions.

Our goal is to construct a renormalized expansion that yields a well-behaved series, without postulating the presence of a Fermi surface. We consider the minimal renormalization scheme where the renormalized self-energy $\Sigma_R[\xi]$ is a quadratic function of the interaction $\xi$. See Fig.~\ref{Fig_quadratic_renormalization} for a Feynman-diagram definition. Let us discuss some possibilities for $\Sigma_{\text{shift}}$. One choice we consider is a BCS-inspired self-energy, which introduces a one-particle gap
\begin{equation}
\Sigma_{\text{R}}=\Sigma_{\text{BCS}}=\frac{\Delta^2}{i\omega_n+\, \gamma \, \xi_{\mathbf{k}}},
\end{equation}
where $\omega_n$ are fermionic Matsubara frequencies, $\xi_{\mathbf{k}}$ is the dispersion of the lattice, and $\Delta$ and $\gamma$ are tunable parameters. Another choice is obtained from the local DMFT self-energy
\begin{equation}
\Sigma_{\text{R}}=\Sigma_{\text{DMFT}}^{\text{loc}}.
\end{equation}

We proceed to numerical results. As a proof of principle we use the bare series as well as the BCS-shifted series to compute the density and potential energy in the Hubbard atom and compare their partial sums to analytically known exact results (Fig.~\ref{Fig_Hubbard_atom}). We observe perfect convergence within $10$ diagram orders of the shifted series whilst the bare series strongly diverges, to the extent that it is impossible to resum. This shortcoming of the bare series is equally true for all further examples that follow.

In Fig.~\ref{Fig_double_occupancy}, we compute the DMFT-shifted series in a strong-coupling, non-perturbative regime of the half-filled Hubbard model, where the bare series fails to converge as we are in the insulating regime. We are able to compute 14 orders of a convergent series that is readily resummated using Pad\'e approximants~\cite{brezinski1996extrapolation, gonnet2013robust, fedor_sigma}.

\begin{figure}
  \includegraphics[width=0.35\textwidth]{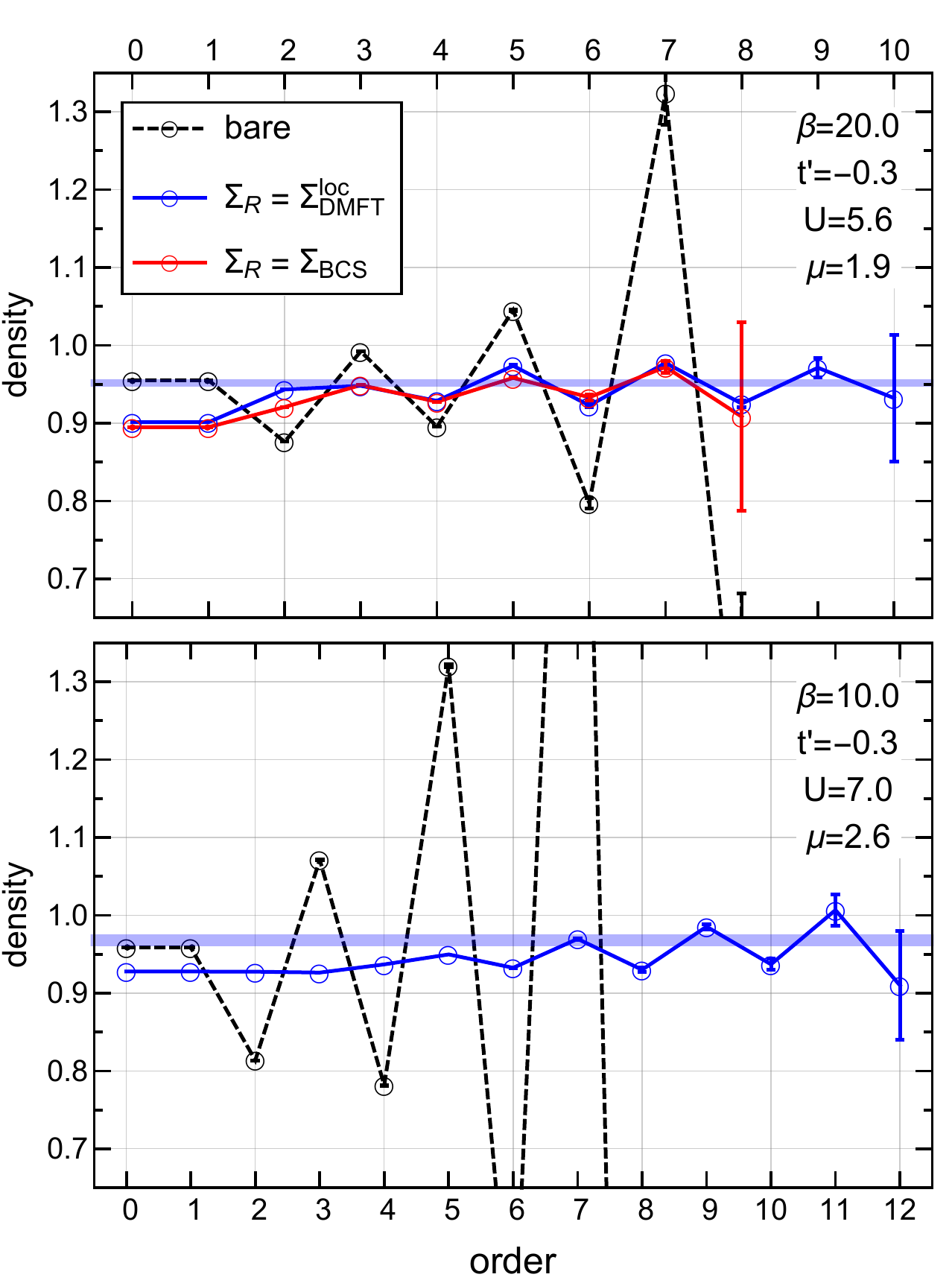}
  \caption{Partial sums for the density in various parameter regimes are shown for the bare series (dashed black lines), the DMFT-shifted series (blue lines), the BCS-type gap shifted series (red line) and compared to the resummated DMFT-shifted series (Pad\'e approximants, horizontal blue line).   \label{Fig_density}}
\end{figure}

In Fig.~\ref{Fig_density}, we show the density at small dopings and in regimes with more immediate relevance to cuprates and pseudogap physics ($t'=-0.3$), at low-temperature and strong interaction. We compute the bare, BCS- and DMFT-shifted series and observe that the last series has the smallest Monte Carlo variance yielding $10-12$ diagrams orders, compared to $8$ orders for the first two. Both shifted series displace a negative-$U$ singularity, associated with a superfluid transition in the attractive Hubbard model~\cite{paiva2004critical}, further away from the origin, thus simplifying the  resummation procedure~\footnote{Unlike the bare expansion results, which is physical for all $U$ (it is  a simple $U$ expansion), generally the renormalized results are physical only for a chosen $U$ point. As our shift is quadratic in $U$, the renormalized expansion will be physical also for $-U$, where the system is in the superfluid s-wave phase at low temperature. This explains the oscillations as a function of the orders seen for both shifts: the system has a singularity at $-U$ for physical reasons, therefore the series cannot converge at $U$.}.

\begin{figure}
  \includegraphics[width=0.45\textwidth]{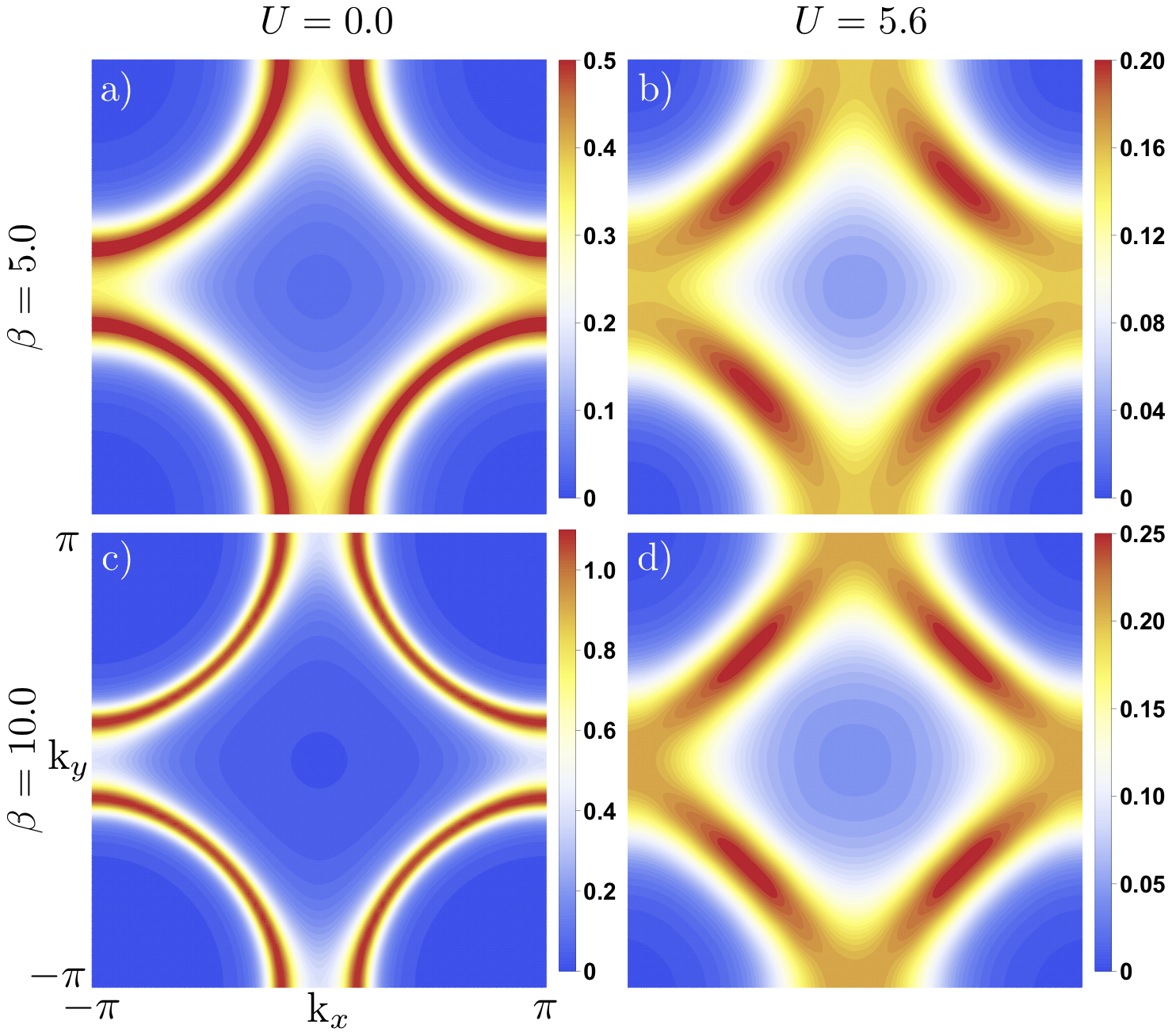}
  \caption{Spectral function of the square-lattice Hubbard model as a function of momentum at imaginary frequency $i\pi/\beta$, density $n=0.950(2)$ (chemical potential $\mu=1.9$), $t^{\prime}=-0.3$, $U=5.6$, and $\beta\in\{5,10\}$. The non-interacting system is shown for comparison.
    \label{Fig_spectral_function}}
\end{figure}

One of the main motivations for this work has been the need to access the non-Fermi liquid regime of the doped square-lattice Hubbard model near half-filling. In Figure~\ref{Fig_spectral_function}, we show the spectral function, computed from a direct sampling of the self-energy at the lowest Matsubara frequency up to order 10 (which is expected to be a good approximation for the zero-frequency spectral function), at inverse temperature $\beta\in\{5,10\}$ and interaction $U \in\{0, 5.6\}$. Order 6 at $\beta=5$ and $U=5.6$ was the limit of the bare computation of Ref.~\cite{wu_controlling}. The spectral function shows significant spectral weight loss near in the antinodal region (near $(\pi, 0)$), while maintaining quasiparticle excitations in the nodal region (near $(\pi/2,\pi/2)$) and differs strongly from the the non-interacting system of same density. The results present strong deviations from the predictions of Fermi-liquid theory, compatible with the phenomenology of the pseudogap regime experimentally found in hole-doped cuprates~\cite{keimer2015pseudogap}. The spectral function shows a remarkable stability as a function of temperature, signaling that the non-Fermi liquid regime is a robust feature of the doped Fermi-Hubbard model, unlike the pseudogap regime found at half-filling~\cite{fedor_hf}.

In conclusion, we have presented a general and efficient way to perform large-order computations for renormalized Feynman diagram series and managed to compute up to 14 expansion orders, roughly double the amount of current state-of-the-art algorithms. We have further shown that it is easy to include non-perturbative information in the contruction of series expansions by using shifted propagators. We have specialized our discussion to renormalization of the one-particle propagator for the 2d Fermi-Hubbard model, where our formalism gave access to non-perturbative regimes at low-temperature and strong coupling, beyond the reach of current numerical techniques, enabling us to illustrate signatures of pseudogap physics in the spectral function. Our formalism can be straightforwardly applied to the renormalization of vertex functions, necessary to access even lower temperatures where superconductivity is expected. The paradigmatic electron gas model, where the renormalization of the Couloumb interaction is necessary to define the model in the thermodynamic limit, could be another important future application. More generally, we believe that our formalism has the potential to yield significant improvements of Feynman diagrammatic computations in quantum chromodynamics, high-energy, solid-state, and statistical physics.

We thank A.~Georges, M.~Koivisto, O.~Parcollet, and T.~Sch\"afer for valuable discussions. This work has been supported by the Simons Foundation within the Many Electron Collaboration framework. DMFT calculations were performed with the CTHYB solver~\cite{seth2016cthyb} of the TRIQS toolbox~\cite{parcollet2015triqs} using HPC resources from GENCI (Grant No. A0070510609). The Flatiron Institute is a division of the Simons Foundation.

\bibliographystyle{ieeetr}
\bibliography{main_biblio}

\end{document}